\documentclass[aps,prb,twocolumn,superscriptaddress]{revtex4}

\usepackage{amsmath,amssymb}
\usepackage{graphicx}
\usepackage{epstopdf}
\usepackage{bm}
\usepackage{color}
\usepackage{mhchem}

\newcommand{\lsco}     {La$_{2-x}$Sr$_x$CuO$_4$}

\newcommand{\lbco}     {La$_{2-x}$Ba$_x$CuO$_4$}

\newcommand{\lmsco}    {La$_{2-x-y}$M$_y$Sr$_x$CuO$_4$}

\newcommand{\tco}      {$T_\text{CO}$}
\newcommand{\tcos}      {$T_\text{CO}^\text{surf}$}
\newcommand{\tso}      {$T_\text{SO}$}
\newcommand{\la}       {$^{139}$La}

\newcommand{\slr}      {$T_1^{-1}$}

\newcommand{\bc}[1]{\textbf{\sffamily #1}}

\begin{document}

\title{Inhomogeneous slowing down of spin fluctuations induced by charge 
stripe order in 1/8-doped lanthanum cuprates} 

\author{S.-H. Baek}
\email[]{sbaek.fu@gmail.com}
%\homepage[]{Your web page}
\affiliation{IFW-Dresden, Institute for Solid State Research,
PF 270116, 01171 Dresden, Germany}
\author{M. H\"ucker}
\affiliation{Condensed Matter Physics and Materials Science Department, 
Brookhaven National Laboratory, Upton, New York 11973, USA} 
\author{A. Erb}
\affiliation{Walther-Mei{\ss}ner-Institut, Bayerische Akademie der Wissenschaften,
Walther-Mei{\ss}ner-Stra{\ss}e 8, D-85748 Garching, Germany}
\author{G. D. Gu}
\affiliation{Condensed Matter Physics and Materials Science Department, 
Brookhaven National Laboratory, Upton, New York 11973, USA} 
\author{B. B\"uchner}
\affiliation{IFW-Dresden, Institute for Solid State Research,
PF 270116, 01171 Dresden, Germany}
\affiliation{Institut f\"ur Festk\"orperphysik, Technische Universit\"at 
Dresden, 01062 Dresden, Germany} 
\author{H.-J. Grafe}
\affiliation{IFW-Dresden, Institute for Solid State Research, PF
270116, 01171 Dresden, Germany}

\date{\today}

\begin{abstract}
%Randomness is an important characteristic of a spin-glass. An example are 
%magnetic ions diluted in a metallic host such  
%as Mn in Cu, the spins of which are randomly oriented below a characteristic temperature.  
%While a spin-glass behavior is mostly caused by time-independent quenched 
%disorder, it is believed that it could also be driven by competing interactions.
%=======
We report \la\ nuclear magnetic resonance (NMR) 
measurements on \lsco\ ($0.07\leq x \leq 0.15$) and  
\lbco\ ($x=1/8$) single crystals, focusing on the spin freezing 
observed in 1/8-doped lanthanum cuprates.  
Charge stripe order seems to induce the inhomogeneous slowing 
down of spin fluctuations toward spin order and compete with 
superconductivity. 
\end{abstract}


\maketitle

Static charge and spin stripe order is a universal characteristic in the lanthanum 
cuprates such as \lbco\ and \lmsco\ 
(M=Nd,Eu) near a hole concentration of $x=1/8$, hereafter called 1/8 
anomaly.\cite{tranquada95,fujita02,hucker07,fink09} 
While static charge stripe order has not been observed in superconducting (SC) \lsco, 
a strong tendency near $x=1/8$ has been 
implicated.\cite{takeya02,park11b} 
Recently, the almost static nature of charge order was proven by 
soft x-ray diffraction measurements,\cite{wu12} which detected
static charge order at \tcos\ = 55 K pinned by small perturbations near 
the surface of LSCO:0.12 single crystal, but not in the bulk of the sample.   
Such a charge ordering tendency and its interplay with 
superconductivity seems to cause a variety of unusual features, such as 
an inhomogeneous SC state\cite{dordevic03,mohottala06} and 
significant effects of magnetic field on static  
antiferromagnetic (AFM) correlations coexisting  
with superconductivity.\cite{katano00, kivelson01, lake02, machtoub05, savici05, chang08,
schafgans10}

Along with these observations, a spin-freezing behavior is a common feature 
observed in lightly-doped 
cuprates.\cite{cho92,chou93,emery93,niedermayer98,julien03,sanna04,ohishi05}
%The CSG state is generally viewed to be caused by randomly localized doped holes, and 
%potentially is related to unidirectional electronic domains or a `nematic' 
%phase.\cite{hanaguri04, kohsaka07} 
While the glassy spin order is rapidly suppressed by increasing doping, it is 
peculiarly enhanced near   
1/8-doping,\cite{savici02,savici05} involving the 
strong enhancement of the NMR spin-lattice relaxation rate,
\slr.\cite{hunt01,simovic03,mitrovic08,baek13a} This unusual reappearance of 
spin order in nearly 1/8-doped LSCO is possibly attributed to the localized 
carriers due to charge ordering \cite{takeya02}.  
This paper presents \la\ \slr\ measurements in  
stripe-ordered LBCO:1/8 as well as superconducting LSCO:1/8. The temperature 
and field dependences of \slr\ suggest that the onset of 
inhomogeneous slowing down of spin fluctuations (SFs) toward spin order 
could be the fingerprint for charge stripe order.

%\section{Sample preparation and experimental details}

The  \lsco\ and \lbco\ were grown with the traveling solvent floating zone 
method, as described in Refs. \onlinecite{baek12a} and 
\onlinecite{gu06}, respectively. 

\la\ NMR measurements were performed on  
\lsco\ single crystals with $x = 0.07$, 0.1, 0.125, and 0.15, and \lbco\ 
single crystal with $x=0.125$, in an external 
field $H$ that ranges from 6 to 16 T, applied along the crystallographic $c$ axis.   
\la\ ($I=7/2$) spin-lattice relaxation rates \slr\ were measured
at the central transition ($+1/2 \leftrightarrow -1/2$) by monitoring 
the recovery of magnetization after saturation with a single $\pi/2$ pulse.  
Then the relaxation data were fitted to the following formula:
\begin{equation}
\begin{split}
\label{eq:T1}
1-\frac{M(t)}{M(\infty)}=
a&\left(\frac{1}{84}e^{-(t/T_1)^\beta}+\frac{3}{44}e^{-(6t/T_1)^\beta}\right.   \\
+ &\left.\frac{75}{364}e^{-(15t/T_1)^\beta}+\frac{1225}{1716}e^{-(28t/T_1)^\beta}\right),
\end{split}
\end{equation}
where $M$ the nuclear magnetization and $a$ a fitting parameter that is
ideally one.  $\beta$ is the stretching exponent, which is less than unity when 
\slr\ becomes spatially distributed due to inhomogeneous spin freezing.  
In Fig. 1(\bc{d}), the typical recovery of $M$ 
versus $t$ and its fit to Eq. (\ref{eq:T1}) are presented for three chosen temperatures 
measured at 10.7 T for LSCO:1/8. 

\begin{figure*}
\centering
\includegraphics[width=0.8\linewidth]{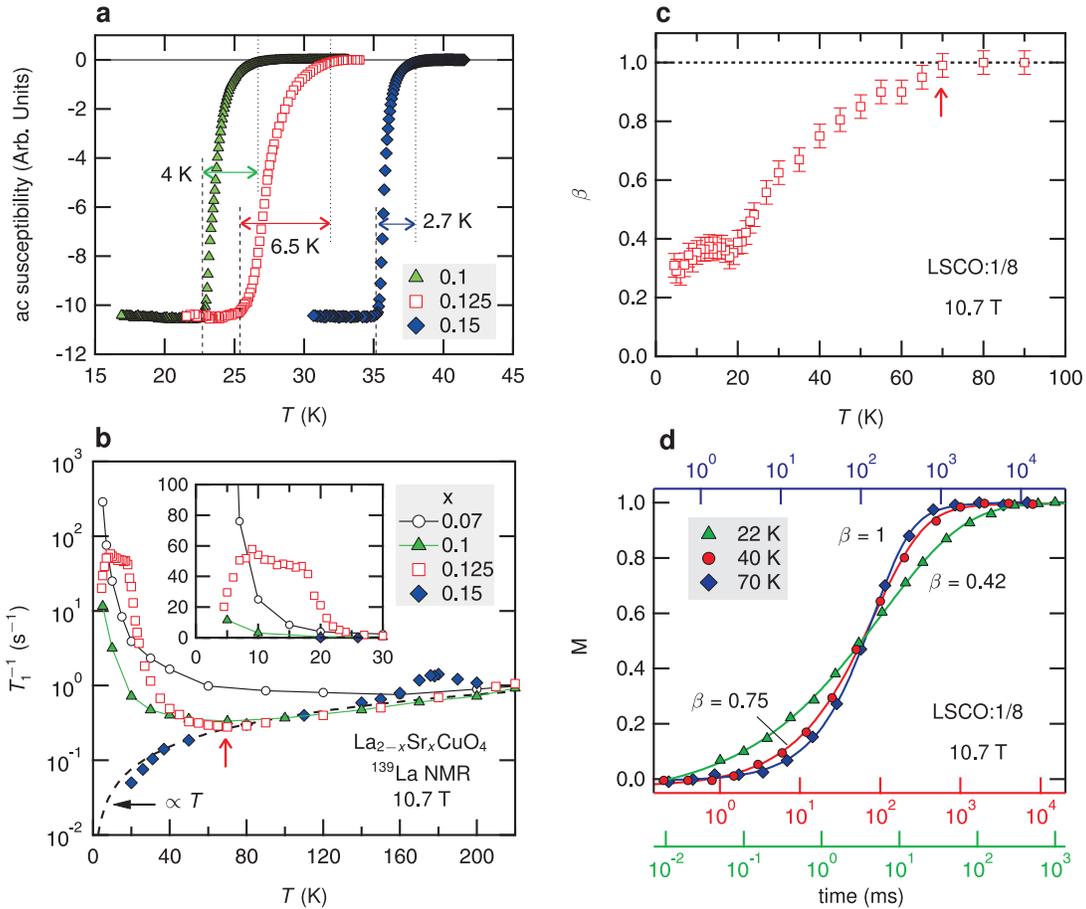}
\caption{\label{fig1} 
(\bc{a}) \textit{In situ} ac susceptibility versus $T$ at three 
dopings measured in the NMR  
circuit in zero field. The SC transition is notably broader at  
$x=1/8$. 
(\bc{b}) \la\ \slr\ versus $T$ as a function  
of $x$ measured at 10.7 T. 
The strong enhancement of \slr\ at $x=0.07$ is drastically suppressed with 
increasing $x$, yielding to
the $T$-linear metallic behavior of \slr\ (denoted  
by dashed curve) at nearly optimal $x=0.15$.
In stark contrast, doping $x=1/8$ causes a strong upturn of \slr, 
which is emphasized on a linear scale in the inset, deviating from the $T$-linear 
behavior at $\sim70$ K (up arrow).  
(\bc{c}) Stretching exponent $\beta$ versus $T$ in LSCO:1/8. The 
deviation of $\beta$ from one occurs near 70 K.  
%There is also a sharp anomaly at $\sim 20$ K.   
(\bc{d}) Recovery of the normalized nuclear magnetization $M$ as 
a function of time $t$ on a semi-log plot at three chosen temperatures. The 
time axis has the same color code as the data.  Solid  
curves are the fits to the data using Eq. (\ref{eq:T1}), yielding $T_1$ and $\beta$.  
To compare the effect of non-unity $\beta$ on the recovery of $M$, the 
maximum of the time axis range for each temperature was set to $10T_1$.  } 
\end{figure*}

Figure 1 (\bc{a}) shows \textit{in situ} ac susceptibility measured in the NMR 
tank circuit in zero 
external field for three compositions of \lsco. Here we identify $T_c$ from the onset of 
the drop (vertical dotted lines), and the resultant values are found to 
be in agreement with SQUID measurements.  
The SC transitions
of the crystals are generally quite sharp, supporting the high quality. 
Nevertheless, we find that the transition for $x=1/8$ is clearly broader 
than for the two  
neighboring dopings $x=0.1$ and 0.15. Similar additional broadening of 
the SC transition near 1/8-doping was previously observed,\cite{kimura96} but 
its origin has rarely been discussed.   
\textit{A priori}, the pronounced broadening in LSCO:1/8 may be related to the 
suppression of $T_c$ due to the strong tendency toward stripe order. Namely, the local 
pinning by the lattice of otherwise slowly fluctuating  
stripe order may cause inhomogeneously distributed $T_c$.  
Indeed, this pinning effect by the lattice accounts for the large reduction 
of $T_c$ observed in nearly 1/8-doped but disordered 
LSCO.\cite{kumagai94,katano00,mitrovic08} 

Figure 1 (\bc{b}) shows the temperature and doping dependence of \la\ \slr\ 
measured at 10.7 T on a semi-log scale. For $x=0.07$, 
\slr\ is enhanced at low $T$ by more than two decades, representing 
the rapid slowing down of SFs toward glassy spin order.\cite{baek12a}
As $x$ is increased, this strong \slr\ enhancement is greatly 
suppressed by more than an order  
of magnitude at $x=0.1$ and disappears completely at nearly optimal doping 
$x=0.15$.\footnote{The 
relatively sharp anomaly of \slr\ near 180 K for $x=0.15$ is   
due to the structural  
phase transition from the high temperature tetragonal (HTT) to the low temperature 
orthorhombic (LTO) phase.}
\begin{figure*}
\centering
\includegraphics[width=0.8\linewidth]{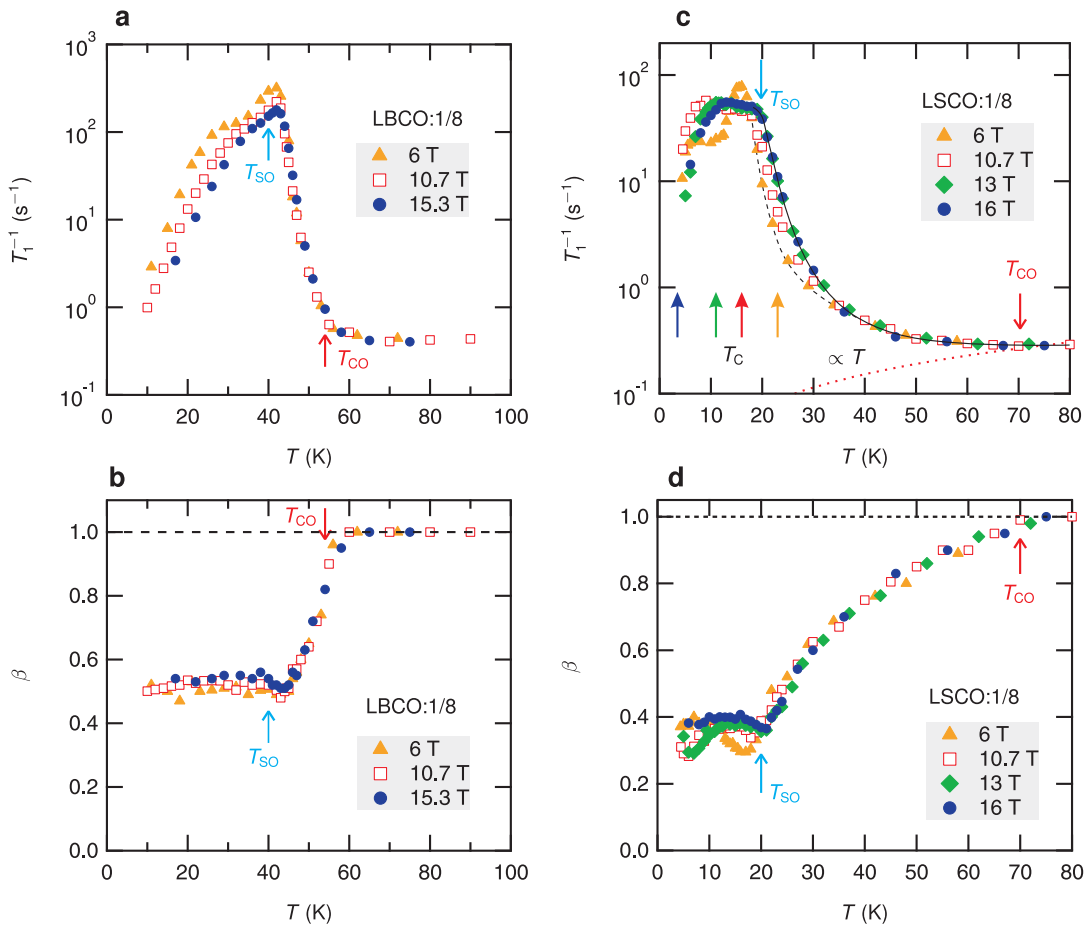}
\caption{\label{fig2} (\bc{a}) \la\ \slr\ (\bc{b}) $\beta$ versus $T$ as a function of 
external field $H$ in LBCO:1/8. Both \slr\ and $\beta$ show the 
abrupt change at near \tco. With 
decreasing $T$, the sharp \slr\ peak centered at $\sim42$ K
is followed by broad  
peak just below $T_\text{SO}$. 
At the same time, $\beta$ reaches a constant below $T_\text{SO}$, essentially 
independent of $H$. 
(\bc{c}) \la\ \slr\ (\bc{d}) $\beta$ versus $T$ as a function of 
$H$ in LSCO:1/8.  
In contrast to LBCO:1/8, \slr\ shows a strong field dependence. In particular, 
the \slr\ upturn is suppressed with decreasing $H$, i.e., with 
increasing $T_c$ which is denoted by the up arrows. 
In the normal state, $\beta (T)$ is almost independent of $H$, as in LBCO:1/8. 
}  
\end{figure*}

Remarkably, 1/8-doping induces an unusual rapid upturn of \slr, which is consistent 
with the enhanced glassy spin order detected in LSCO:0.12 by  
muon spin rotation ($\mu$SR)\cite{savici02,savici05} and 
NMR.\cite{hunt01,simovic03,mitrovic08} 
Deviating at $\sim70$ K with respect to the linear $T$ dependence which may be 
expected to be followed in this doping regime, \slr\ 
rises sharply until it bends over at $\sim 18$ K.  
Instead of forming a sharp local maximum expected in a  
conventional spin-glass phase, however, \slr\ continues to increase before it 
drops abruptly at $\sim 8$ K. 
The stretching exponent $\beta$ from Eq. (\ref{eq:T1}) also starts to 
deviate from unity near 70 K, as shown in Fig. 1(\bc{c}).
A $\beta$ value less than unity indicates a spatial  
distribution of \slr\ and, therefore, can be used as a measure for
magnetic inhomogeneity of the spin system. 
Thus our \slr\ shows that SFs are inhomogeneously slowed 
down below $\sim 70$ K. 

At near 1/8-doping, the doped holes are expected to be largely delocalized,\cite{kohsaka07} 
yielding the metallic behavior as was confirmed for $x=0.15$. 
In this case, since quenched disorder is shown not to be responsible for the glassy 
behavior in LSCO:1/8,\cite{mitrovic08} the entity that 
drives the unusual spin freezing near 1/8-doping is likely related to the 
1/8-anomaly.  Specifically, together with the unusual  
broadening of the SC transition shown in Fig. 1(\bc{a}), we conjecture that 
charge stripe order, although it may be still rapidly fluctuating on the NMR time 
scale ($\sim\mu$s), may generate the randomness (e.g., localized holes) that 
could inhomogeneously slow down the spin fluctuations.  

In order to check whether the inhomogeneous slowing down and charge order are related, we 
performed similar measurements in stripe-ordered LBCO:1/8, which are 
presented in Fig. \ref{fig2} (\bc{a}).   
The \slr\ peak reveals a strongly asymmetric peak whose height depends on the 
external field (i.e., the  
resonance frequency $\omega_n=\gamma_n H$ where $\gamma_n$ is the nuclear 
gyromagnetic ratio). The field dependence of \slr\ clearly 
shows that the high  
temperature side of the peak is frequency-independent.   
This low temperature frequency dependence of the \slr\ peak is   
qualitatively understood by the Bloembergen, Purcell, and 
Pound (BPP) model\cite{bloembergen48} which is appropriate for describing the 
continuous slowing down of SFs,\cite{suh00,curro00,simovic03}
\begin{equation}
\label{eq:bpp}
     T_1^{-1} = \langle\gamma_n^2 h_\perp^2\rangle \frac{\tau_c}{1+\omega_n^2\tau_c^2},
\end{equation}
where $h_\perp$ the local field fluctuating at the nuclear site, and the 
electron correlation time $\tau_c$  
is in general given by $\tau_c=\tau_\infty \exp(E_a/T)$ with $E_a$ the 
activation energy. 

The BPP model predicts that the high temperature side of the \slr\ peak 
is frequency independent, while the peak height decreases with increasing 
field.  This is consistent with the main features of the \slr\ peak in Fig. 
\ref{fig2} (\bc{a}). The 
similar BPP behavior is also observed in  
another stripe-ordered LESCO:0.13.\cite{baek13a}
Most importantly, both \slr\ and $\beta$ manifest a very sharp anomaly just above 
\tco, indicating that charge stripe order\cite{hucker11} triggers the inhomogeneous 
slowing down. 
Another surprise is that \textit{below the spin 
ordering temperature} $T_\text{SO}$,\cite{hucker11}
\slr\ falls off much slower than above \tso.   
At the same time, $\beta$ is almost saturated to a constant
regardless of the external field strength, which 
could be interpreted to reflect stabilized spin order. 

Returning to LSCO:1/8, the fact that the onset of both the \slr\ enhancement 
and the deviation of $\beta$ from unity is much less clear than LBCO:1/8 could 
reflect the rapidly fluctuating or significantly disordered nature of charge 
order in LSCO:1/8. Nevertheless, the  
onset temperature could be identified with reasonable certainty, suggesting that
charge order seemingly occurs at  
\tco\ = 70(10) K, which appears to be independent of a magnetic field, as 
would be expected for  
charge stripe order above $T_c$.\cite{hucker13,chang12,blackburn13,blanco-canosa13}
At low temperatures, on the other hand,  
when superconductivity is nearly quenched at 16 T, the temperature 
dependence of $\beta$ is very similar to that 
of LBCO:1/8, as shown in Fig. 2(\bc{d}). In particular, it becomes almost  
a constant just below a sharp anomaly at 20 K.  The similar $T$-dependence 
of $\beta$ in the two materials suggests that \tso\ in LSCO:1/8 as well as in LBCO:1/8 
represents a true phase transition, rather than a progressive crossover, to spin order, 
despite the strong glassy character.  

Taking it for granted that the inhomogeneous slowing down of SFs for 1/8-doped 
La cuprates is induced by charge ordering, 
NMR might further probe the interplay between  
stripe order and superconductivity in LSCO:1/8, which is a much better 
superconductor than its Ba doped relative. 
In fact, as shown in Fig. 2(\bc{c}) and 
(\bc{d}), the field dependence of \slr\ appears to reveal such an interplay.   
At high fields ($\ge 13$ T), i.e. when 
superconductivity is   
sufficiently suppressed, the high temperature side of the \slr\ peak is 
independent of $H$, which is similar to LBCO:1/8 and conform with the standard 
BPP model.
However, the \slr\ upturn clearly becomes suppressed with decreasing  
field, i.e. increasing $T_c$. 
%That means that the standard BPP model breaks down when  
%superconductivity becomes strong. 
This breakdown of the BPP behavior is indicative of a competition between  
charge order and superconductivity. 
An obvious question is then why the reduction of the \slr\ 
upturn in Fig. 2(\bc{c}) 
occurs well above the bulk $T_c (H)$. We think that this behavior is consistent with 
two-dimensional (2D) SC correlations\cite{li07, schafgans10, mosqueira11, 
bilbro11, stegen13} which are known to  
coexist with charge order above the bulk $T_c$.\cite{berg09a}
This idea is substantiated by the fact that the temperature at which 
the \slr\ upturn is  
suppressed seems to be limited to the bulk $T_c\sim 32$ K in zero field,
implying that a magnetic field frustrates interlayer coupling but 
preserves intralayer coupling.\cite{berg09a}    

While the slowing down of SFs above \tso\ provides 
information regarding the charge order and its competing relationship with 
superconductivity, 
the complex field dependence that appears below \tso\ \textit{in the SC state} for
both \slr\ and $\beta$ does not permit us to reach a conclusion about the 
relationship between spin and SC orders. 
%Nevertheless, we note that the \slr\ data 
%at 6 T as shown in Fig. 2(\bc{c}) are qualitatively similar to those of 
%LBCO:1/8, i.e. the relatively  
%distinctive \slr\ peak just above \tso\ is followed by a broad peak  
%at lower temperatures. This suggests that the considerable 
%suppression of the broad \slr\ peak in  
%the low-$T$ region for LSCO:1/8 corresponds to the suppression of spin order due to 
%superconductivity. On the other hand, 
Nevertheless, the saturated $\beta$ below \tso\ at 16 T 
suggests that spin order is further stabilized as superconductivity is 
weakened. 

In conclusion, we reported \la\ \slr\ measurements in \lsco\ 
($0.07\leq x \leq 0.15$) and \lbco\ ($x=1/8$). Our data suggest that charge 
ordering may trigger inhomogeneous slowing down of spin fluctuations toward spin 
stripe order. On the basis of our NMR results, we propose that charge ordering 
may set in at 70(10) K and  
compete with superconductivity in La$_{1.875}$Sr$_{0.125}$CuO$_4$.

%\section*{acknowledgment}

We thank J. Geck for useful and delightful discussion. This work has been 
supported by the DFG through FOR 538 (Grants No. BU887/4 and No. ER342/1-3) and 
SPP1458 (Grant No. GR3330/2). S.B. acknowledges support by the DFG 
Research Grant No. BA 4927/1-1. M.H. acknowledges support by the U.S. 
Department of Energy (DOE), under Contract No. DE-AC02-98CH10886.

\bibliography{mybib}

\end{document}